\documentstyle[preprint, revtex]{aps}
\def\beq{\begin{eqnarray}}
\def\eed{\end{eqnarray}}
\begin{document}
\draft

\begin{title}
\begin{center}
{Theory of Photoemission from the Copper Oxide Material}
\end{center}
\end{title}
\author{Shiping Feng$^{1,2,3}$, Yun Song$^{1,2}$}
\begin{instit}
$^{1}$CCAST (World Laboratory) P. O. Box 8730, Beijing 100080, China and \\
$^{2*}$Department of Physics, Beijing Normal University, Beijing
 100875, China and \\
$^{3}$National Laboratory of Superconductivity, Academia Sinica,
 Beijing 100080, China \\
\end{instit}
\begin{abstract}
A mean-field theory which satisfying the electron on-site local constraint
in the relevant regime of density for the high temperature superconductors
is developed. Within this approach, the electron spectral function, the
electron dispersion, and the electron density of states of copper oxide
materials are discussed, and the results are qualitative consistent with
the numerical simulations.
\end{abstract}

The discovery of the high temperature superconductivity in copper oxide
materials has initiated an enormous theoretical effort on quantum
antiferromagnets in two-dimensions (2D) \cite{n1,n2,n3}. Although
experiments have not yet isolated the key elements of the electronic
structure necessary for a global understanding of the physical
properties of copper oxide materials, a significant body of reliable
and reproducible data has been amassed by using many probes \cite{n2,n3}.
The normal-state properties of these materials exhibit a number of
anomalous properties in sense that they do not fit in the conventional
Fermi-liquid theory \cite{n4}. When undoped, with one hole per copper site,
these materials are antiferromagnetic Mott insulators \cite{n2}. When
doped with sufficient carriers, they are superconductors with no the
antiferromagnetic long-range order \cite{n2}. There is strong evidence
from transport and neutron scattering that no sharp phase transition
occurs, and the strong correlations are very important to the electronic
structure \cite{n5}. The angle-resolved photoemission spectroscopy is the
definitive way to determine the electron dispersion relation for any
materials. In copper oxide materials, the angle-resolved photoemission
experiments \cite{n6} have produced interesting data that introduce
important constraints on theories. It has been shown \cite{n2,n3} that
the hole-doped copper oxide materials exhibit universal properties
likely induced by the behavior of carriers in their common copper oxide
sheets. It is believed \cite{n1} that the physics of these materials may
be effective described by a 2D, large U, single-band Hubbard model. In
the large U limit, the Hubbard model is transfered into $t$-$J$ model
acting on the space with no doubly occupied sites. Furthermore, Zhang
and Rice \cite{n1} derived the $t$-$J$ model from a multiband Hubbard
model described 2D copper oxide planes.

It is clearly very important to establish appropriate formalism for
the problem and to show that this leads to behavior similar to that
seen in experiments. In order to account for real experiments under
the $t$-$J$ model, the crucial requirement is to impose the electron
local constraint \cite{n7}. The local nature of the constraint is of
prime important, and its violation may lead to some unphysical results
\cite{n8}. Recently a fermion-spin theory based on the charge-spin
separation is proposed \cite{n9} to incorporate this constraint. In this
approach, the electron on-site local constraint for single occupancy is
satisfied even in the mean-field approximation (MFA). In the framework
of the fermion-spin theory, the ground-state properties, such as, the
ground-state energy, ground-state kinetic energy, phase separation,
specific heat data, and doping dependence of the antiferromagnetic
long-range order, are discussed \cite{n9,n10,n11} and the results are in
qualitative agreement with experiments and numerical simulations. In
this paper, we develope a mean-field theory in optimal doping regime within
the fermion-spin theory to study the photoemission spectrum, the electron
dispersion, and the electron density of states, which is useful for
understanding of the electronic structure of the copper oxide materials.

We start from the $t$-$J$ model which describes the electrons moving
on a planar square lattice,
\begin{eqnarray}
H = -t\sum_{\langle ij\rangle\sigma}C^{\dagger}_{i\sigma}C_{j\sigma} + h.c.
- \mu \sum_{i\sigma}C^{\dagger}_{i\sigma}C_{i\sigma} +
J\sum_{\langle ij\rangle}{\bf S}_{i}\cdot {\bf S}_{j} ,
\end{eqnarray}
where $C^{\dagger}_{i\sigma}$ ($C_{i\sigma}$) are the electron creation
(annihilation) operators,
${\bf S}_{i}=C^{\dagger}_{i}{\bf \sigma} C_{i}/ 2$ are spin operators
with ${\bf \sigma}=(\sigma_{x},\sigma_{y},\sigma_{z})$ as Pauli
matrices, and $\mu$ is the chemical potential. The summation
$\langle ij\rangle$ is carried over nearest nonrepeated bonds. The
$t$-$J$ Hamiltonian (1) is supplemented by the on-site local constraint,
$\sum_{\sigma}C^{\dagger}_{i\sigma}C_{i\sigma}\leq 1$, {\it i.e.},
there be no doubly occupied sites. With the help of the fermion-spin
transformation \cite{n9}
\begin{eqnarray}
C_{i\uparrow}=h^{\dagger}_{i}S^{-}_{i},~~~~
C_{i\downarrow}=h^{\dagger}_{i}S^{+}_{i},
\end{eqnarray}
where the spinless fermion operator $h_{i}$ keeping track of the charge
(holon) while the pseudospin operator $S_{i}$ keeping track of the spin
(spinon), the $t$-$J$ model (1) can be rewritten as \cite{n9}
\begin{eqnarray}
H = -t\sum_{\langle ij\rangle}h_{i}h^{\dagger}_{j}(S^{+}_{i}S^{-}_{j}
+S^{-}_{i}S^{+}_{j}) +  h. c.  \nonumber \\
- \mu \sum_{i}h^{\dagger}_{i}h_{i} +
J\sum_{\langle ij\rangle}(h_{i}h^{\dagger}_{i})({\bf S}_{i}
\cdot {\bf S}_{j})(h_{j}h^{\dagger}_{j}) ,
\end{eqnarray}
where $S^{+}_{i}$ and $S^{-}_{i}$ are pseudospin raising and lowering
operators, respectively. It is shown \cite{n9} that the constrained
electron operator can be mapped exactly onto the fermion-spin
transformation defined with an additional projection operator.
However, this projection operator is cumbersome to handle for the
actual calculation possible in 2D, we have dropped it in
Eq. (3). It has been shown in Ref. \cite{n9} that such treatment leads
the errors of the order $\delta$ in counting the number of spin states,
which is negligible for small doping $\delta$. Within the MFA, the
$t$-$J$ model (3) can be decoupled as,
\begin{mathletters}
\begin{eqnarray}
H=H_{t}+H_{J}-8Nt\chi\phi, ~~~~~~~~~~~~~~~~~~~~~~~~~~~ \\
H_{t}=2\chi t\sum_{i,\eta}h^{\dagger}_{i+\eta}h_{i}
-\mu \sum_{i}h^{\dagger}_{i}h_{i}, ~~~~~~~~~~~~~~~~~~~~~~~~~~~~~~~ \\
H_{J}={1\over 2}J_{eff}\epsilon \sum_{i,\eta}(S^{+}_{i}S^{-}_{i+\eta}
+S^{-}_{i}S^{+}_{i+\eta})+
J_{eff} \sum_{i,\eta}S^{z}_{i}S^{z}_{i+\eta},
\end{eqnarray}
\end{mathletters}
with $\eta=\pm \hat{x},\pm \hat{y}$, $N$ is the number of sites, and
$J_{eff}=J[(1-\delta)^{2}-\phi ^{2}]$. The nearest-neighbor spin
bond-order amplitude $\chi$ and holon particle-hole parameter
$\phi$ are defined as $\chi =\langle S^{+}_{i}S^{-}_{i+\eta}\rangle$
and $\phi =\langle h^{\dagger}_{i}h_{i+\eta}\rangle$, respectively,
where the site subscripts of the order parameters $\chi$ and $\phi$
have been dropped since the system is translation invariant. In this
mean-field level, the spinon part is described by an anisotropic
Heisenberg model with the anisotropic parameter is given by,
\begin{eqnarray}
\epsilon = {J_{eff} +2t\phi \over J_{eff}}.
\end{eqnarray}

The quantum spin operators obey the Pauli spin algebra, and this problem
can be discussed in terms of the two-time spin Green's function within
the Tyablikov scheme \cite{n111}. In this case, the one-particle spinon
and holon two-time Green's functions are defined as,
\begin{mathletters}
\begin{eqnarray}
D(i-j,t-t')=-i\theta(t-t')\langle [S^{+}_{i}(t),S^{-}_{j}(t')]\rangle
=\langle\langle S^{+}_{i}(t);S^{-}_{j}(t')\rangle\rangle , \\
D_{z}(i-j,t-t')=-i\theta(t-t')\langle [S^{z}_{i}(t),S^{z}_{j}(t')]\rangle
=\langle\langle S^{z}_{i}(t);S^{z}_{j}(t')\rangle\rangle ,
\end{eqnarray}
\end{mathletters}
and
\begin{eqnarray}
g(i-j,\tau-\tau')=-i\theta(t-t')\langle [h_{i}(t),h^{\dagger}_{j}(t')]
\rangle =\langle\langle h_{i}(t);h^{\dagger}_{j}(t') \rangle\rangle ,
\end{eqnarray}
respectively, where $\langle \ldots \rangle$ is an average over the
ensemble. Because the spinon system is an anisotropic, we have defined
the two spinon Green's function $D(i-j,t-t')$ and $D_{z}(i-j,t-t')$ to
describe the spinon propagations. The time-Fourier transform of the
two-time Green's function satisfies the equation,
\begin{eqnarray}
\omega\langle\langle A;B\rangle\rangle ={1\over 2\pi}
\langle [A,B]\rangle + \langle\langle [A,H];B \rangle\rangle ,
\end{eqnarray}
therefore the Green's functions can be obtained by applying
the Tyablikov decoupling technique, and the correlation functions
can be obtained by the spectral representations as
\begin{eqnarray}
\langle B(t')A(t)\rangle\ =i\int^{\infty}_{-\infty}{d\omega\over 2\pi}
{\langle\langle A;B \rangle\rangle_{\omega+i0^{+}}-\langle\langle A;B
\rangle\rangle_{\omega-i0^{+}}\over e^{\beta\omega}-1}
e^{-i\omega(t-t')}.
\end{eqnarray}

Recently, we \cite{n9,n10,n11} have employed the fermion-spin theory
to study the ground-state properties of the 2D $t$-$J$ model and obtained
some interesting results. Within the random-phase approximation, we
\cite{n11} have shown that the antiferromagnetic long-range order is
destroyed by hole doping of the order $\sim 5\%$ for the reasonable value
of the parameters $t/J=5$. Then in the following discussions, we only
study the systems in the optimal doping regime ($20\% >\delta >5\%$),
where there is no the antiferromagnetic long-range order, {\it i.e.},
$\langle S^{z}_{i}\rangle =0$. In this case, the basic equations for
the mean-field spinon two-time Green's function in one-dimension have
been discussed in detail by Kondo and Yamaji \cite{n12}. Following their
discussions, we can obtain the mean-field spinon Green's functions of
$H_{J}$ in 2D,
\begin{mathletters}
\begin{eqnarray}
D({\bf k},\omega)={\Delta [(2\epsilon\chi_{z}+\chi)\gamma_{k}-
(\epsilon\chi+2\chi_{z})]\over 2\omega (k)}\left (
{1\over \omega -\omega (k)}-{1\over \omega +\omega (k)}\right ) , \\
D_{z}({\bf k},\omega)={\Delta\epsilon\chi (\gamma_{k}-1)\over
2\omega_{z}(k)}\left ({1\over \omega -\omega_{z}(k)}-{1\over \omega
+\omega_{z}(k)}\right ) , ~~~~~~~~~~~~~~~
\end{eqnarray}
\end{mathletters}
where
$\gamma_{{\bf k}}={1\over Z}\sum_{\eta}e^{i{\bf k}\cdot\hat{\eta}}$,
and
\begin{mathletters}
\begin{eqnarray}
\omega^{2}(k)=\Delta^{2}\left (\alpha\epsilon(\chi_{z}\gamma_{k}-
{1\over Z}\chi )(\epsilon\gamma_{k}-1)+[\alpha C_{z}+{1\over 4Z}
(1-\alpha)](1-\epsilon\gamma_{k})\right ) \nonumber \\
+\Delta^{2}\left ({1\over 2}\alpha\epsilon\chi\gamma_{k}(\gamma_{k}
-\epsilon) + {1\over 2}\epsilon[\alpha C+{1\over 2Z}
(1-\alpha)](\epsilon-\gamma_{k}) \right ), \\
\omega^{2}_{z}(k)=\Delta^{2}\left (\epsilon^{2}[\alpha C+{1\over 2Z}
(1-\alpha)]-\alpha\epsilon\chi\gamma_{k}-{1\over Z}\alpha\epsilon\chi
\right )(1-\gamma_{k}),
\end{eqnarray}
\end{mathletters}
with $\Delta =2ZJ_{eff}$, $Z$ is the number of the nearest neighbor
sites, and the order parameters
$\chi_{z}=\langle S^{z}_{i}S^{z}_{i+\eta}\rangle$,
$C={1\over Z^{2}}\sum_{\eta,\eta'}\langle S^{+}_{i+\eta}S^{-}_{i+\eta'}
\rangle$, and $C_{z}={1\over Z^{2}}\sum_{\eta,\eta'}\langle
S^{z}_{i+\eta}S^{z}_{i+\eta'}\rangle$. In order not to violate the
sum rule of the correlation function $\langle S^{+}_{i}S^{-}_{i}
\rangle={1\over 2}$ in the case $\langle S^{z}_{i}\rangle =0$, the
important decoupling parameter $\alpha$ has been introduced as these
discussed by Kondo and Yamaji \cite{n12}, which can be regarded as the
vertex corrections. At half-filling, the $t$-$J$ model is reduced as
the isotropic antiferromagnetic Heisenberg model, where $\epsilon=1$,
$\chi_{z}={1\over 2}\chi$, and $C_{z}={1\over 2}C$ in the rotational
symmetrical case, and we obtain
$D_{z}({\bf k},\omega)={1\over 2}D({\bf k},\omega)$, which is
just these discussed by Shimahara and Takada \cite{n13}.

The mean-field Green's function of $H_{t}$ for holons is very
simple, and can be written as
\begin{eqnarray}
g({\bf k},\omega)={1\over \omega-(\varepsilon_{k}-\mu)},
\end{eqnarray}
where $\varepsilon_{k}=2Z\chi t\gamma_{k}$. With help of the spinon and
holon Green's functions and the spectral representations of the
correlation functions, the order parameters $\chi$, $C$, $\chi_{z}$,
$C_{z}$, $\phi$, and chemical potential $\mu$ can be obtained by
the self-consistent equations,
\begin{mathletters}
\begin{eqnarray}
\chi = {1\over N}\sum_{k}\gamma_{k}{\Delta [(2\epsilon\chi_{z}+\chi)
\gamma_{k}-(\epsilon\chi+2\chi_{z})]\over 2\omega (k)}{\rm coth}
({\beta\omega (k)\over 2}), ~~~~\\
C ={1\over N}\sum_{k}\gamma^{2}_{k}{\Delta [(2\epsilon\chi_{z}+\chi)
\gamma_{k}-(\epsilon\chi+2\chi_{z})]\over 2\omega (k)}{\rm coth}
({\beta\omega (k)\over 2}),~~\\
{1\over 2} ={1\over N}\sum_{k}{\Delta [(2\epsilon\chi_{z}+\chi)
\gamma_{k}-(\epsilon\chi+2\chi_{z})]\over 2\omega (k)}{\rm coth}
({\beta\omega (k)\over 2}),~~~~~~~~~~~~\\
\chi_{z} ={1\over N}\sum_{k}\gamma_{k}{\Delta\epsilon\chi
(\gamma_{k}-1)\over 2\omega_{z}(k)}{\rm coth}({\beta\omega_{z}(k)
\over 2}),~~~~~~~~~~~~~~~~~\\
C_{z} ={1\over N}\sum_{k}\gamma^{2}_{k}{\Delta
\epsilon\chi (\gamma_{k}-1)\over2\omega_{z}(k)}{\rm coth}
({\beta\omega_{z}(k)\over 2}), ~~~~~~~~~~~~~~~~~~~~ \\
\phi ={1\over 2N}\sum_{k}\gamma_{k}\left (1-{\rm th}{\beta (\varepsilon_{k}
-\mu)\over 2}\right ),~~~~~~~~~~~~~~~~~~~~~~~~~ \\
\delta = {1\over 2N}\sum_{k}
\left (1-{\rm th}{\beta (\varepsilon_{k}
-\mu)\over 2}\right ).~~~~~~~~~~~~~~~~~~~~~~~~~~~~~~~~
\end{eqnarray}
\end{mathletters}
As we have shown in the previous works \cite{n11} that the present MFA
self-consistent calculation is just the usual self-consistent
Hartree-Fock approximation.

We \cite{n14} have performed a numerical calculation for these mean-field
self-consistent equations. The results for the order parameters at optimal
doping regime are very close to our previous works \cite{n9,n10} based on
the 2D Jordan-Wigner approach, the detailed discussions will be given
elsewhere \cite{n14}. In this paper, we hope to discuss the electronic
structure of copper oxide materials, and therefore it need to calculate
the electron Green's function $G(i-j,t-t')=\langle\langle C_{i\sigma}(t);
C^{\dagger}_{j\sigma}(t')\rangle\rangle$. According the fermion-spin
transformation (2), the electron Green's function is a convolution of the
spinon Green's function $D(i-j,t-t')$ and holon Green's function
$g(i-j,t-t')$, and can be obtained at the mean-field level as,
\begin{eqnarray}
G({\bf k},\omega )={1\over N}\sum_{p}{\Delta [(2\epsilon\chi_{z}+\chi)
\gamma_{p}-(\epsilon\chi+2\chi_{z})]\over 2\omega (p)}\times \nonumber\\
\left ({F_{1}({\bf k},{\bf p})\over \omega -\omega (p)+\varepsilon_{p-k}}
+{F_{2}({\bf k},{\bf p})\over \omega +\omega (p)+\varepsilon_{p-k}}
\right )~~,
\end{eqnarray}
where $F_{1}({\bf k},{\bf p})=n_{B}(\omega_{p})+
n_{F}(\varepsilon_{p-k}-\mu)$, $F_{2}({\bf k},{\bf p})=1+
n_{B}(\omega_{p})-n_{F}(\varepsilon_{p-k}-\mu)$, with
$n_{B}(\omega_{p})$ and $n_{F}(\varepsilon_{p-k}-\mu)$ are the boson and
fermion distribution functions for spinons and holons, respectively.
From the electron Green's function (14), we obtain the electron spectrum
function,
\begin{eqnarray}
A({\bf k},\omega )=-2{\rm Im}G({\bf k},\omega)=2\pi {1\over N}
\sum_{p}{\Delta [(2\epsilon\chi_{z}+\chi)\gamma_{p}
-(\epsilon\chi+2\chi_{z})]\over 2\omega(p)}\times \nonumber \\
\left ( F_{1}({\bf k},{\bf p})\delta (\omega -\omega (p)+
\varepsilon_{p-k})+F_{2}({\bf k},{\bf p})\delta (\omega
+\omega (p)+\varepsilon_{p-k}) \right )~~~ .
\end{eqnarray}
In the $t$-$J$ model, the doubly occupied Hilbert space has been pushed
to infinity as Hubbard $U\rightarrow \infty$ and therefore the spectrum
function only describes the lower Hubbard band. Our mean-field result of
the spectral functions at the doping $\delta =0.12$ for the parameter
$t/J=2.5$ is shown in Fig. 1 (solid line). For comparison, the exact
diagonalization and quantum Monte Carlo result at $\delta =0.12$ for
$t/J=2.5$ obtained by Moreo {\it et al.} \cite{n15} is also shown in
Fig. 1 (dashed line). Although the particular details of the spectral
function and dispersion may differ from compound to compound, some
qualitative features seem common to all copper oxide materials. Hence a
quantitative comparison between theory and experiment is still early, but
the qualitative tendency of the spectral function and dispersion in an
adequate theoretical description should be consistent with experiments
and numerical simulations. In the present mean-field theory, the most
important feature is that the intensity peaks is qualitative consistent
with the numerical simulation \cite{n15}. The low energy peak is well
defined at all momenta, and the positions of the dominant peaks in
$A({\bf k},\omega)$ as a function of momentum are shown in Fig. 2, which
is also in qualitative agreement with the numerical simulation \cite{n15}
and the experimental result \cite{n6}.

Now we consider the electron density of states, which is defined as
\begin{eqnarray}
\rho (\omega )={1\over N} \sum_{k}A({\bf k},\omega ).
\end{eqnarray}
The numerical analysis of the electron density of states in the $t$-$J$
model as a function of doping has been done by many authors \cite{n16}.
On the other hand, oxygen x-ray absorption spectra measured \cite {n17} on
$La_{2-x}Sr_{x}CuO_{4}$ may be interpreted in terms of a picture in which
hole doping introduces carriers into the lower hand. Our mean-field result
at doping $\delta =0.12$ and $\delta=0.06$ for the parameter $t/J=2.5$ is
shown in Fig. 3. We find that the chemical potential $\mu$ moves from
nearly zero at small doping $\delta=0.06$ to the top edge of the lower
Hubbard band, which also is qualitative consistent with the numerical
simulation \cite{n16} and experimental result \cite{n17}.

In summary, we have developed a mean-field theory which satisfying the
electron on-site local constraint in the relevant regime of density for
the high temperature superconductors, namely in the vicinity of optimal
doping within the fermion-spin theory. Within this mean-field theory, we
have study the electron spectral function, the electron dispersion, and
the electron density of states of copper oxide materials, the results are
qualitative consistent with the numerical simulations and experiments.

At zero doping, the $t$-$J$ model reduces to the antiferromagnetic
Heisenberg model, which has a local SU(2) symmetry in the fermion
representation \cite{n19}. This symmetry does imply that the spinon
particle (hole) state with spin up is the same state as a spinon hole
(particle) with spin down. In the conventional slave-boson theory
\cite{n20}, the SU(2) is broken to U(1) upon doping, and this U(1) gauge
degree of freedom is introduced to incorporate with the single occupancy
local constraint. However, in the fermion-spin theory, the SU(2)
symmetry is also broken upon doping. Moreover, since the local
constraint is satisfied exactly even in the MFA,
the extra gauge degree of freedom occurring in the slave-particle
approach does not appear in the here, which is consistent with these
discussions in Ref. \cite{n21}.

Final we also note that recently Wen and Lee \cite{n22} have developed
a slave-boson theory for the $t$-$J$ model at underdoped regime which
preserves SU(2) symmetry, they argued that spin gap phase at the
underdoped regime can be understood as the staggered flux phase.
However, the area of the Fermi surface produced by this SU(2)
mean-field theory is larger than the predicted by the Luttinger
theorem which reveals a drawback of the SU(2) mean-field theory
at the optimal doping, and therefore the U(1) mean-field theory
is better at the optimal doping.

\acknowledgments
The authors would like to thank Mr. Z. B. Huang and Prof. Z. X. Zhao
for helpful
discussions. This work is supported by the National Science Foundation
Grant No. 19474007 and the Trans-Century Training Programme Foundation
for the Talents by the State Education Commission of China.

\references{

\bibitem [*] {add} Mailing address.

\bibitem {n1} P. W. Anderson, in "{\it Frontiers and Borderlines
in Many particle Physics}", edited by R. A. Broglia and J. R.
Schrieffer (North-Holland, Amsterdam, 1987)p. 1; Science {\bf 235},
1196 (1987); F. C. Zhang and T. M. Rice, Phys. Rev. B {\bf 37},
3759 (1988).

\bibitem {n2} A. P. Kampf, Phys. Rep. {\bf 249}, 219 (1994), and
references therein.

\bibitem {n3} See, {\it e.g.}, "{\it High Temperature
Superconductivity}", Proc. Los Alamos Symp., 1989, K. S. Bedell,
D. Coffey, D. E. Meltzer, D. Pines, and J. R. Schrieffer, eds.
( Addison-Wesley, Redwood City, California, 1990); E. Manousakis,
Rev. Mod. Phys. {\bf 63}, 1 (1991), and references therein.

\bibitem {n4} See, e.g., the review, L. Yu, in {\it Recent Progress
in Many-Body Theories, edited by T. L. Ainsworth et al.} (Plenum,
New York, 1992), Vol. 3, p. 157.

\bibitem {n5} N. W. Preyer {\it et al.}, Phys. Rev. B{\bf 39}, 11563
(1989); J. P. Falck {\it et al.}, Phys. Rev. B{\bf 48}, 4043 (1993);
B. Keimer {\it er al.}, Phys. Rev. B{\bf 46}, 14034 (1992).

\bibitem {n6} D. S. Dessau {\it et al.}, Phys. Rev. Lett. {\bf 71},
2781 (1993); B. O. Wells {\it et al.}, Phys. Rev. Lett. {\bf 74},
964 (1995); A. A. Abrikosov {\it et al.}, Physica C{\bf 214}, 73
(1993); D. M. King {\it et al.}, Phys. Rev. Lett. {\bf 73}, 3298
(1994); K. Gofron {\it et al.}, Phys. Rev. Lett. {\bf 73}, 3302
91994).

\bibitem {n7} T. K. Lee and Shiping Feng, Phys. Rev. B {\bf 38},
11809 (1988); S. Liang, B. Doucot, and P. W. Anderson, Phys. Rev. Lett.
{\bf 61}, 365 (1988); Z. Liu and E. Manousakis, Phys. Rev. B {\bf 40},
11437 (1989); D. A. Huse and V. Elser, Phys. Rev. Lett. {\bf 60}, 2531
(1988).

\bibitem {n8} Shiping Feng, J. B. Wu, Z. B. Su, and L. Yu,
Phys. Rev. B{\bf 47}, 15192 (1993); L. Zhang, J. K. Jain, and
V. J. Emery, Phys. Rev. B{\bf 47}, 3368 (1993).

\bibitem {n9} Shiping Feng, Z. B. Su, and L. Yu, Phys. Rev. B
{\bf 49}, 2368 (1994); Mod. Phys. Lett. B{\bf 7}, 1013 (1993);
Shiping Feng {\it et al.}, in"{\it Superconductivity and Strongly
Correlated Electron Systems}", p124, C. Noce, A. Romano, and G.
Scarpetta, eds. (World Scientific, 1994).

\bibitem {n10} Shiping Feng, Physica C{\bf 232}, 119 (1994); X. Xu,
Y. Song, and Shiping Feng, Mod. Phys. Lett. B{\bf 9}, 1623 (1995);
X. Xu, Y. Song, and Shiping Feng, Acta Physica Sinica, {\bf 45},
1390 (1996).

\bibitem {n11} Shiping Feng, Phys. Rev. B{\bf 53}, 11671 (1996); Shiping
Feng, Y. Song, and Z. B. Huang (unpublished).

\bibitem {n111} S. V. Tyablikov, {\it Methods in the Quantum Theory of
Magnetism} (Plenum, New York, 1967).

\bibitem {n12} J. Kondo and K. Yamaji, Prog. Theor. Phys. {\bf 47},
807 (1972).

\bibitem {n13} H. Shimahara and S. Takada, J. Phys. Soc. Jpn. {\bf 60},
2394 (1991).

\bibitem {n14} Yun Song and Shiping Feng (unpublished).

\bibitem {n15} A. Moreo {\it ea al.}, Phys. Rev. B {\bf 51}, 12045
(1995).

\bibitem {n16} E. Dagotto {\it et al.}, Phys. Rev. B{\bf 45}, 10741
(1992).

\bibitem {n17} C. T. Chen {\it et al.}, Phys. Rev. Lett. {\bf 66},
104 (1991); H. Romberg {\it et al.}, Phys. Rev. B {\bf 42}, 8768
(1990).

\bibitem {n19} I. Affleck {\it et al.}, Phys. Rev. B {\bf 38}, 745
(1988); E. Dogotto {\it et al.}, Phys. Rev. B {\bf 38}, 2926 (1988).

\bibitem {n20} G. Kotliar and J. Liu, Phys. Rev. B {\bf 38}, 5142
(1988); Y. Suzumura {\it et al.}, J. Phys. Soc. Jpn. {\bf 57},
2768 (1988).

\bibitem {n21} F. C. Zhang {\it et al.}, Supercond. Sci. Technol.
{\bf 1}, 36 (1988).

\bibitem {n22} X. G. Wen and P. A. Lee, Phys. Rev. Lett. {\bf 76},
503 (1996).

}

\figure{Spectral function $A({\bf k},\omega )$ of the 2D $t$-$J$ model
within the fermion-spin mean-field theory (solid line) and the exact
diagonalization and quantum Monte Carlo methods (dashed line) for the
parameter $t/J=2.5$ }

\figure{Position of the the dominant peaks in $A({\bf k},\omega )$ as
a function of momentum.}

\figure{Electron spectral density of the 2D $t$-$J$ model at the
parameter $t/J=2.5$ for (a) doping $\delta=0.12$ and (b) $\delta=0.06$.}

\end{document}